\documentclass[letterpaper]{article}
\usepackage{aaai}
\usepackage{times}
\usepackage{helvet}
\usepackage{courier}
\usepackage{graphicx}
\usepackage{multicol}
\usepackage{multirow}
\usepackage{bbm}
\usepackage{subfig}
\usepackage{balance}
\usepackage{amsmath,amssymb}
\usepackage{booktabs}
\usepackage{placeins}
\usepackage[font=small,labelfont=bf]{caption}
\newtheorem{hyp}{H}
\newcommand\Mark[1]{\textsuperscript#1}

\makeatletter
\renewcommand{\paragraph}{%
  \@startsection{paragraph}{4}%
  {\z@}{1.5ex \@plus 1ex \@minus .2ex}{-1em}%
  {\normalfont\normalsize\bfseries}%
}
\makeatother

\begin{document}
\title{Keep Your Friends Close and Your Facebook Friends Closer: \\A Multiplex Network Approach to the Analysis of Offline and Online Social Ties} 
\author{Desislava Hristova\Mark{1}, Mirco Musolesi\Mark{2} and Cecilia Mascolo\Mark{1}\\
\Mark{1} Computer Laboratory, University of Cambridge, UK\\
\Mark{2} School of Computer Science, University of Birmingham, UK}
\maketitle

\begin{abstract}
Social media allow for an unprecedented amount of interaction between people online.
A fundamental aspect of human social behavior, however, is the tendency of people to associate themselves with like-minded individuals, forming homogeneous social circles both online and offline. In this work, we apply a new model that allows us to distinguish between social ties of varying strength, and to observe evidence of homophily with regards to politics, music, health, residential sector \& year in college, within the online and offline social network of 74 college students. We present a multiplex network approach to social tie strength, here applied to mobile communication data - calls, text messages, and co-location, allowing us to dimensionally identify relationships by considering the number of communication channels utilized between students. We find that strong social ties are characterized by maximal use of communication channels, while weak ties by minimal use. We are able to identify 75\% of close friendships, 90\% of weaker ties, and 90\% of Facebook friendships as compared to reported ground truth. We then show that stronger ties exhibit greater profile similarity than weaker ones. Apart from high homogeneity in social circles with respect to political and health aspects, we observe strong homophily driven by music, residential sector and year in college. Despite Facebook friendship being highly dependent on residence and year, exposure to less homogeneous content can be found in the online rather than the offline social circles of students, most notably in political and music aspects. 
\end{abstract}

\section{Introduction}

Despite the vast potential for communication through social media such as Facebook, users tend to interact mostly with their closest friends~\cite{backs2011}. Friends tend to come from similar socio-demographic backgrounds, share common interests and information. This presents evidence of homophily, or the long-standing social truth that ``similarity begets friendship"~\cite{plato1968}. While it has interesting implications in social networks in terms of link prediction~\cite{aiello2012}, resilience~\cite{newman2003}, and preferential attachment~\cite{papa2012}, homophily also leads to the localization of information and resources into socio-demographic space~\cite{mcpherson2001}.
Conversely, diversity in social contacts has been shown to be of great importance for social and economic wellbeing, both at individual and community levels~\cite{Eagle2010}. Access to diverse information and resources in social networks can result in easier access to jobs and opportunities. Much of the diversity in information that we experience both offline and online come from weak ties ~\cite{gran1983,bakshy2012}. 

Social media sites like Facebook have had a profound effect on the way we maintain close and distant social relationships, on their number and their diversity, and the cultivation of our social capital~\cite{vitak12,vitak14}. With more weak ties online we have access to more diverse news, opinions and information in general. It increasingly appears that while homophily is strongly present in traditional social networks offline~\cite{mcpherson2001}, there is an emergence of ``heterophily" online, where people are exposed to and engage with information mainly from others who are dissimilar~\cite{bakshy2012,mac2011}. \emph{While in light of social diversity this is a sought effect, no comparison between online and offline social circles has confirmed this until now.}

In this work we introduce a new multiplex measure of social tie strength and apply it to uncover the presence of homophily within a community of 74 students, with respect to political orientation, music preferences, health habits, and situational factors. We then measure whether students can potentially be exposed to more diversity in these categories through their offline or online (Facebook) social circles. Our methodology is inspired by Fischer's early work on homophily~\cite{fischer1982}, where it was observed that the more types of relations that exist between two people (e.g., friends, kin, neighbor), the stronger their bond, and the stronger the effects of homophily (similarity) between them. Although very different from kinship relations, the same principle of tie strength and depth was found in media multiplexity with regards to various types of media in the organizational environment~\cite{haythornthwaite1998}, where stronger ties interact through more types of media than weaker ones. In this work, we use this concept in student interactions and apply the resulting multiplex social tie measure to homophily, all the while considering both online and offline relationships in the community. 

The contributions made in this work can be summarized as follows:

\begin{itemize}
\item We propose a novel methodology for studying online and offline interactions based on communication multiplexity. We apply it to the multi-channel communication of a group of 74 MIT co-resident students, taking into account phone calls, text message, and co-location data. We show that the greater number of channels utilized  between two individuals (multiplexity), the higher the probability of a strong social tie such as close friendship. 

\item Considering the pairwise similarity of user profiles based on political orientation, music preference, health indicators, and situational factors such as residential sector and year in college, we show that there is a significant positive relationship between communication multiplexity and profile similarity. 

\item We distinguish between the effects of homogeneity and homophily on a multiplex network level, across categories. We observe that while the political and health similarity between students can be credited to the overall homogeneity within the community, there is distinct evidence of homophily based on music and situational factors.

\item Finally, by observing the difference in composite similarity between one's offline and online social circles, we find that \emph{online interactions are more diverse} in terms of exposure to more varied musical preferences and political opinions. However, diversity online is relative to situational factors offline such as commonality in residence and year in college and these appear to be driving factors for Facebook friendship.
\end{itemize}

The remainder of this work is structured as follows: we first present a review of the literature on multiplexity, offline and online homophily and social capital, followed by a description of our research hypotheses, methodology and data. We then present a quantitative evaluation of our hypotheses followed by a discussion of the implications of the relative diversity found online, and the role of social media such as Facebook in diversifying human interactions.

\section{Related Work}

\paragraph{Multiplexity and Tie Strength.} Multiplexity has been explored in a wide range of systems from global air-transportation \cite{cardillo2013} to massive online multiplayer games~\cite{Szell19072010}. A comprehensive review of multiplex (multilayer) network models can be found in~\cite{kivela2013}. While multiplexity in most systems creates additional complexity such as layer interdependence, in social networks multiplexity can be used to define the strength of a tie. 

In~\cite{hay2005}, the author studied the implications of media usage on social ties in an academic organization and discovered the same dependency that Fischer observed in the 80's in rural areas~\cite{fischer1982} - that multiplex ties indicate a stronger bond.  Typically measured through the use of a single media, in this work we consider multiplexity as a measure for tie strength.
 
\paragraph{Online and Offline Homophily.} Recent research on homophily [\emph{lit.} love of same], used sensors for tracking mobility and interactions \emph{offline} in the same college setting as this present work, and showed that physical exercise, residential sector, and on-campus activities are the most important factors for the formation of social relationships, placing emphasis on spatio-temporal activities~\cite{pentland2011a}. Furthermore, dynamic homophily based on political opinion was studied in the same context during the 2008 US presidential election by means of Bluetooth scanning: the researchers observed increased proximity around the presidential debates between students with the same political orientation~\cite{pentland2011b}. 

Social network research has further focused on the effects of homophily expressed in interactions \emph{online}, where findings suggest that most of the content shared comes from weak and diverse ties. The authors in~\cite{bakshy2012} showed that in the context of Facebook, while strong ties are consistently more influential, weak ties are collectively more important and users consume and share information produced largely by those with whom they interact infrequently. Diversity in online social network exchanges has also been observed in Twitter, where users re-tweet more content from topically dissimilar ties~\cite{mac2011}. In our work, we link social tie strength to evidence of homophily across political, music, health and situational categories.

\paragraph{Social Capital.} Homophily often creates localization of resources and information, leading to a decrease in potential social capital. Social capital is the value embedded in social networks in the form of contacts who can potentially offer support (strong ties) and opportunities (weak ties), also convertible to other types of capital~\cite{burke2013}. The usage of Facebook and other social media has been associated with increased social capital with comparison to non-users~\cite{lampe2013,steinfield2009}. In the educational setting, social capital is also increased with the usage of social media, specifically bringing benefits of social adjustment, support and persistence~\cite{Gray2013193}. In this work, we do not study the effects of online media usage on social capital previously explored in~\cite{wellman2001} but rather compare between the online and offline social circles of students and observe the differences in terms of potential diversity across categories.

\section{Analyzing Offline \& Online Social Ties}

We study homogeneity and homophily in the community as it relates to multiplex tie strength. Ultimately our aim is to discern whether students are exposed to greater variety of opinions and information in their online or offline social circles. We define ties and their respective strength from mobile communication. Based on the theory of media multiplexity ~\cite{hay2005}, which states that strongly tied pairs make use of a greater number of available online media, we hypothesize that a similar relationship exists in communication measured through mobile phones (calls, SMS, and co-location data):

\begin{hyp}
The more communication channels utilized, the stronger the tie.
\end{hyp}

We consider politics, music, health habits, residential sector \& year in college as diversity factors. We use the term \emph{diversity} in this study to define the variety of information introduced by ties as opposed to \emph{homogeneity}. Based on Fischer's observation that greater multiplexity results in greater similarity~\cite{fischer1982}, we hypothesize that on both the edge and neighborhood network levels:

\begin{hyp}
The more communication channels utilized, the greater the similarity observed with respect to politics, health, music and residence \& year in college.
\end{hyp}

In order to confirm the presence of homophily, we study the multiplex communication network weighted considering the number of types of interactions and we measure the weighted distance between individuals in the resulting graph. We test whether:

\begin{hyp}
The closer two nodes are in the weighted network of interactions, the greater their similarity with respect to politics, health, music and residence \& year in college.
\end{hyp}

In light of recent studies of homophily online, and in particular findings, which suggest that there exists a negative homophily phenomenon online~\cite{bakshy2012,mac2011}, we hypothesize that: 

\begin{hyp}
Online social circles are more diversified than offline social circles with respect to politics, health, music and residence \& year in university.
\end{hyp}

We will next formulate our multilayer approach, later applied to gain understanding of the multiplexity in the communication network and the social ties between pairs. 

\section{Multiplex Graph Model \\ For Social Interaction Analysis}

In this section, we present our model for analyzing social networks where interactions happen through different communication channels. We firstly introduce the concept of multiplex graphs and then define the social interaction aggregations used in the remainder of this work.

\paragraph{Multiplex graphs.} Different types of interactions and relationships between the same two nodes can be successfully captured by means of the abstraction of a multiplex (multilayer) graph. We define a \textit{multiplex graph} $\cal{M}$ as an ensemble of $M$ graphs, each corresponding to an interaction type, and therefore representing a layer in the multilayer (multiplex) graph. 
We indicate the $\alpha$-th layer of the multiplex as $G^\alpha(V^\alpha, E^\alpha)$. Therefore, we can denote the sequences of graphs composing the $M$-layer graph as:

\begin{equation}
{\cal{M}}= \{G^1(V^1,E^1),...,
G^\alpha(V^\alpha,E^\alpha),...,G^M(V^M,E^M)\}
\end{equation}

An adjacency matrix $A^\alpha$ is associated with each graph $G^\alpha(V^\alpha, E^\alpha)$ representing the layer $\alpha$ of the multiplex. Therefore, $\cal{M}$ can also be described with a sequence of adjacency matrices $A = [A^1, ..., A^\alpha, ... ,A^M]$. We denote by $a^\alpha_{ij}$ the element of the matrix $A^\alpha$ at layer $\alpha$ representing the link between nodes $i$ and $j$ on that layer. 

\begin{figure}[t!]
\centering
\includegraphics[scale=0.5]{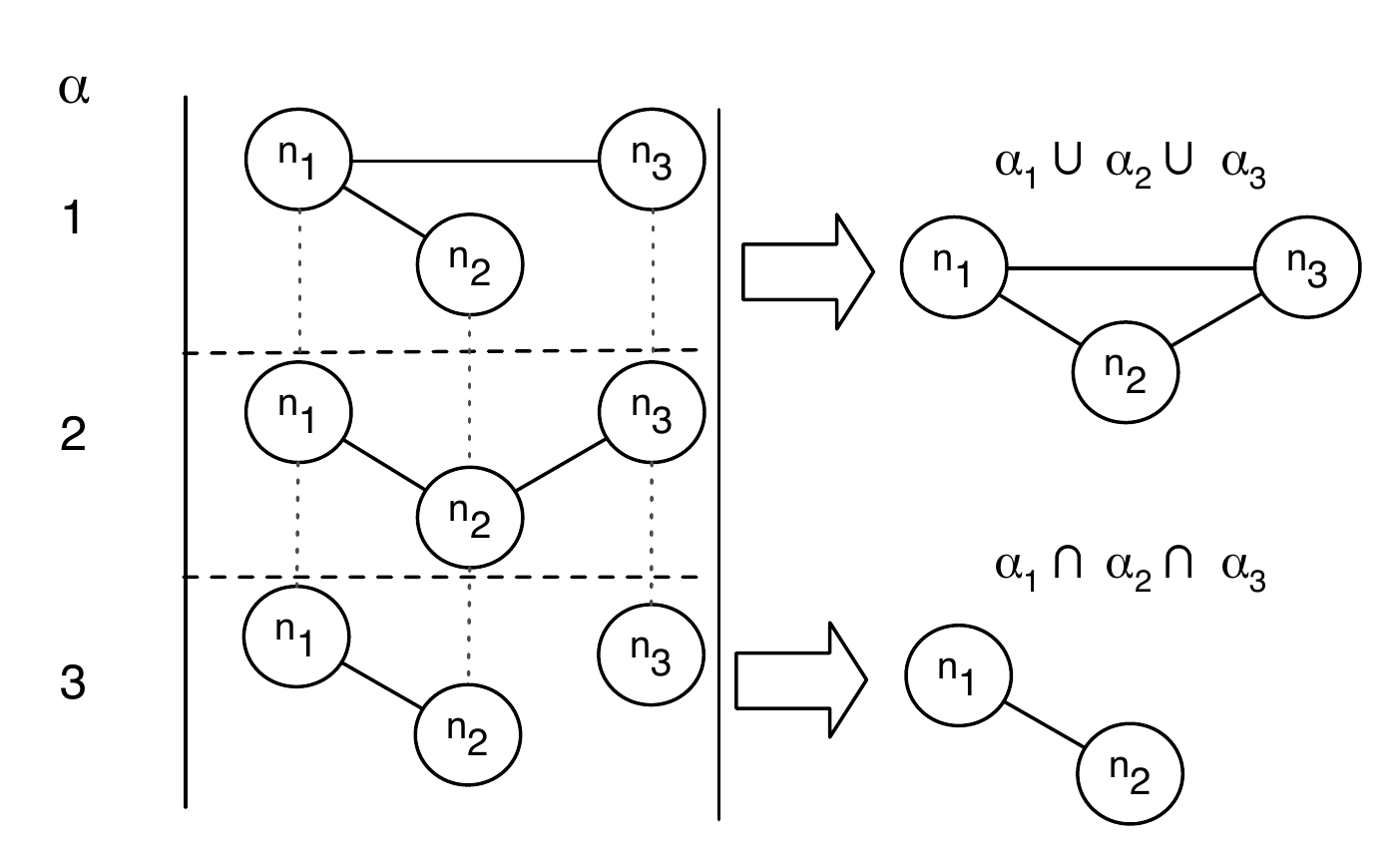}
\caption{Union and Intersection aggregations are shown on the right, where $\alpha$ indicates the layer; the original networks/layers of the multiplex are shown on the left.}
\label{fig:model}
\end{figure}

\paragraph{Union \& intersection aggregated graphs.} A multiplex graph of the social interactions of individuals can be represented through different layer aggregations to better understand the properties of their tie, such as its strength. In order to define these aggregations we need to first define the concepts of union and intersection of two graphs.

We define the union $G^\alpha \bigcup G^\beta$ of two graphs $G^\alpha$ and $G^\beta$ represented respectively by the adjacency matrices $A^\alpha$ and $A^\beta$ as the graph described by the adjacency matrix $A^{\alpha \cup \beta}$ with elements $a^{\alpha \cup \beta} = 1$ if $a^\alpha _{ij}=1$ or $a^\beta _{ij} = 1$, 0 otherwise.

We define the intersection $G^\alpha \bigcap G^\beta$ of two graphs $G^\alpha$ and $G^\beta$ represented respectively by the adjacency matrices $A^\alpha$ and $A^\beta$ as the graph described by the adjacency matrix $A^{\alpha \cap \beta}$ with elements $a^{\alpha \cap \beta} = 1$ if $a^\alpha _{ij}=1$ and $a^\beta _{ij} = 1$, 0 otherwise.

Given these definitions, we will consider the following two types of layer aggregations of the multiplex $\cal{M}$:
\emph{the union graph} $\bigcup_{\cal{M}}$
defined as

\begin{equation}
\bigcup_{\cal{M}} = G^1 \bigcup G^2 \ldots \bigcup G^M
\end{equation}

i.e., the graph aggregation in which an edge between two nodes is present if it is present in \textit{at least} one layer; and the \emph{intersection graph} $\bigcap_{\cal{M}}$ defined as:

\begin{equation}
\bigcap_{\cal{M}} = G^1 \bigcap G^2 \ldots \bigcap G^M
\end{equation}

i.e., the graph aggregation in which an edge between two nodes is present if it is present in \textit{all} the layers. Both the union and intersection aggregations are illustrated in Fig.~\ref{fig:model}.

It is worth noting that it is possible to restrict this aggregation to a subset of graphs $\{ \alpha, \beta, \gamma, ...\}$ and define for example the union graph over the set of layers $\{ \alpha, \beta, \gamma, \ldots\}$ as the graph $\bigcup_{\cal{M,}\{ \alpha, \beta, \gamma, \ldots\}}$ corresponding to the union of the graphs of layers $\alpha, \beta, \gamma, \ldots$. The intersection graph aggregation over a set of layers can be defined in a similar way.


\section{Dataset}
The open-access MIT Social Evolution dataset\footnote{http://realitycommons.media.mit.edu/socialevolution.html} contains details of the everyday life of a group of students between October 2008 and May 2009 \cite{pentland2012}. These students co-reside in two adjacent college residence buildings during term time. Details of their health habits, political orientation, music preferences, social relationships (online and offline), and mobile communication were collected during this period, allowing for a rich analysis of the relationships between their characteristics, social ties and communication.

\subsection{Communication Layers}
The multiplex interaction graph is built by combining different communication layers.
Three types of interactions can be extracted from the mobile phone data - physical proximity data (whether pairs of users were within 10 meters of each other, inferred from Bluetooth); phone call record data (who called whom); and SMS data (who texted whom). Each of these communication layers is a network in its own right (Table \ref{table:nets}).

\begin{table}[h!]
\caption{Network specifications. \emph{We take into account only interactions between students, ignoring external ones in the dataset.}} 
\small
\centering
\begin{tabular}{l l c c c c c }
\hline \hline
network & type
& avg degree & nodes & edges\\
\hline
calls & directed
& 5.8 & 69 & 401\\
\hline
SMS & directed
& 2.12 & 33 & 70\\
\hline
proximity & undirected
& 61.2 & 74 & 4,526\\
\hline
\end{tabular}
\label{table:nets}
\end{table}

\begin{figure}[t!]
\centering {\label{fig:deg1}\includegraphics[scale=0.3]{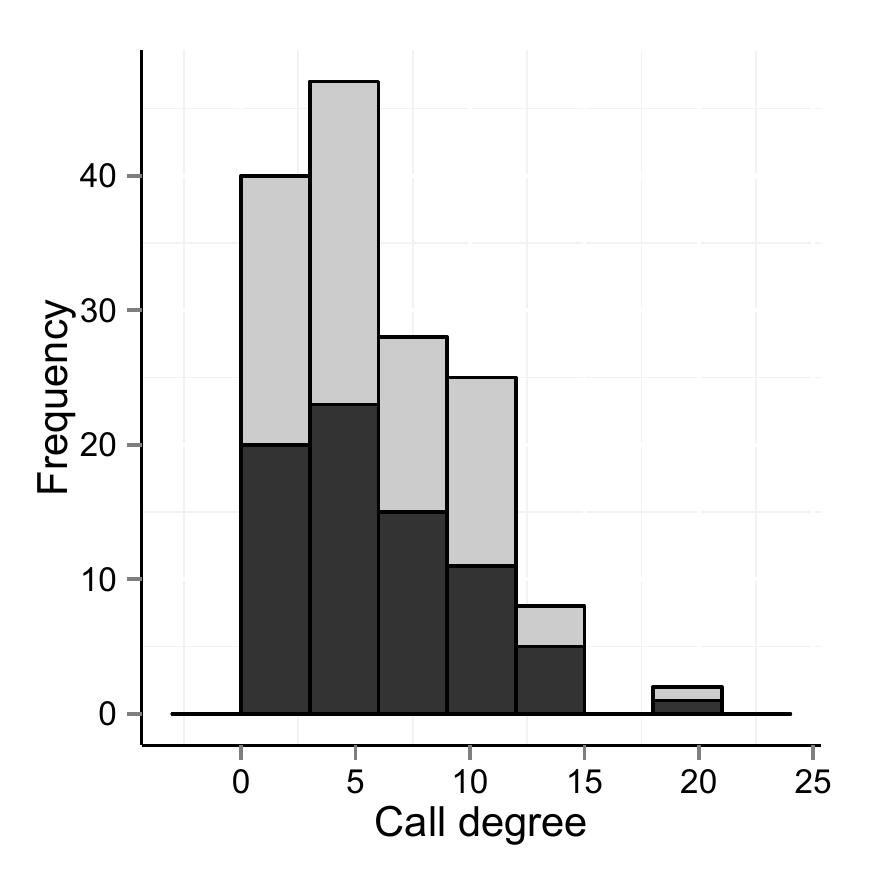}}
{\label{fig:deg2}\includegraphics[scale=0.3]{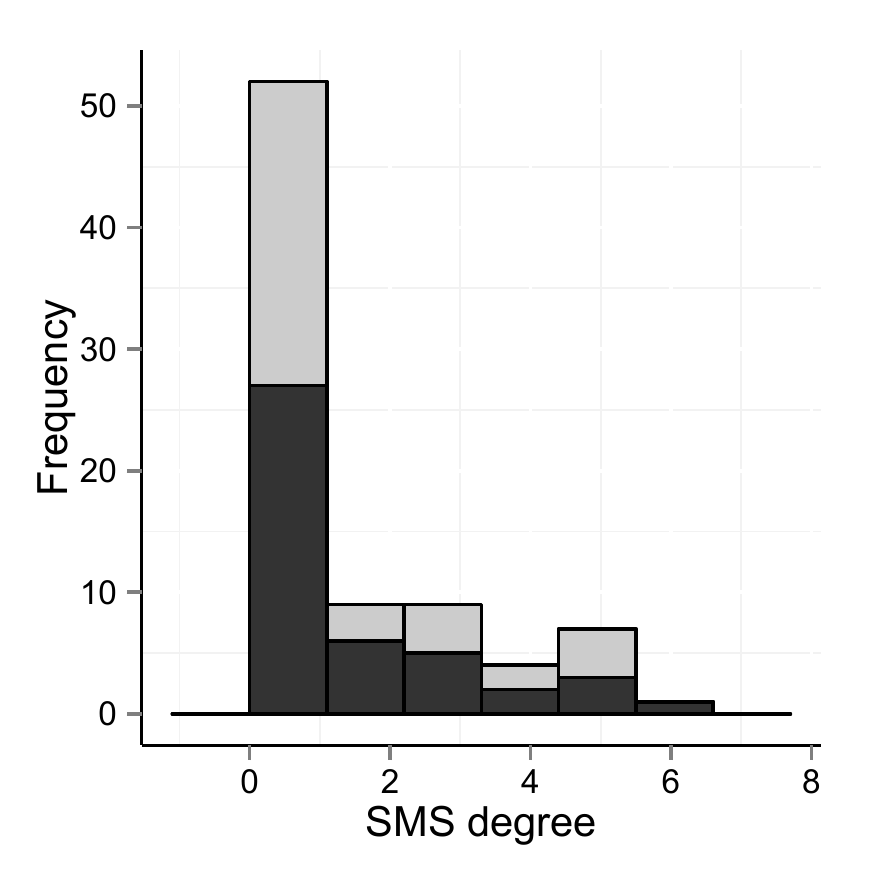}}
{\label{fig:deg3}\includegraphics[scale=0.3]{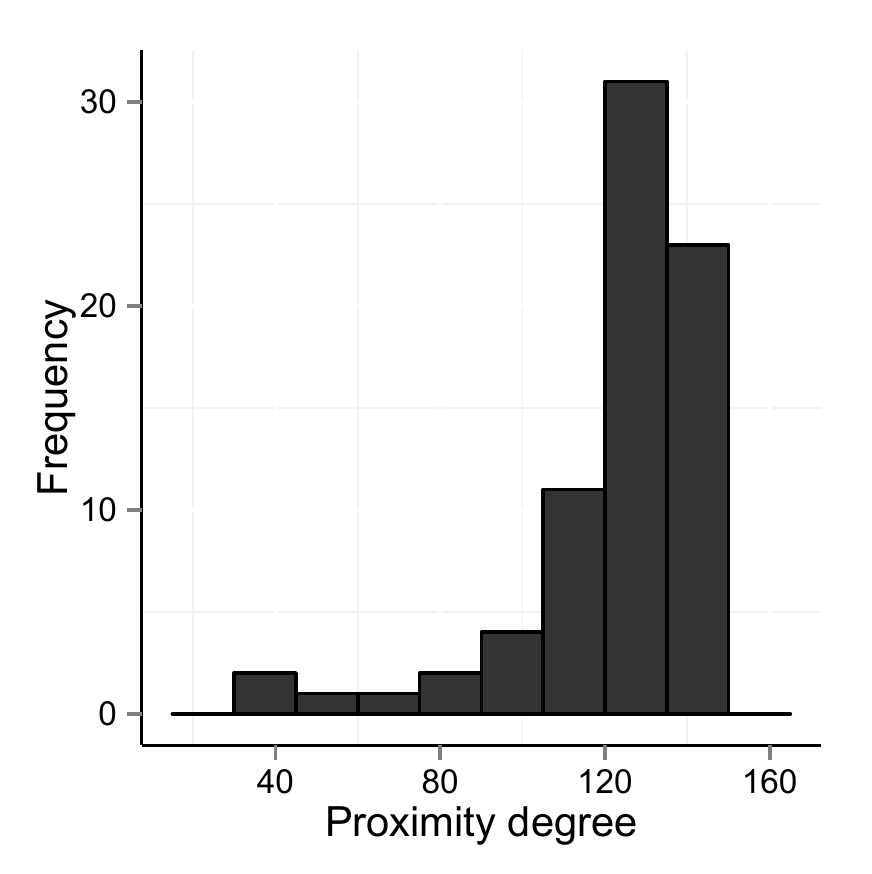}}
\caption{Degree distributions for the call, SMS and proximity networks. In-degree is in light, while out-degree is in dark (proximity is undirected).}
\label{fig:degs}
\end{figure}

The degrees across the networks are uncorrelated and the degree distributions show that each communication layer is utilized to a different extent and purpose (Fig.~\ref{fig:degs}). While the proximity layer has a high average degree, the other two layers have a low average degree in comparison, giving us initial insight into the social dynamics of the student community - many students meet many others but few talk to many on the phone or text. The three layers complement each other, and in combination represent three basic communication channels of human interactions - passing time together, talking on the phone and sending messages. 

If we denote the set of edges in the proximity layer as $P$, the call layer as $C$ and the SMS layer as $S$, the relationship between the three communication layers can be described as $S\subset C\subset P$, meaning that all participants have been co-located with another participant and are part of the proximity layer but not all have called or sent a text message to another participant. This is because the proximity layer is prevalent, and the basis of all further communication when a connection exists between two nodes, likely due to the fact that all students are co-residing and possibly have lectures together. All pairs with a call edge or a SMS edge also have a proximity edge. Incidentally, almost all pairs with a SMS edge also have a call edge (92\% overlap), which may not be generalizable to the case of other communication networks. Overall, the density of the student network is such that 83\% of all nodes are connected on at least one layer - the proximity layer.

\subsection{Social Relationships}

Social relationships reported by the participants form the ground truth for our social tie analysis. Details of data collection methodology are described in~\cite{pentland2012}. We consider three types of reported social relationships: \emph{Facebook friendship}, \emph{Socializing twice per week}, and \emph{Close friendship}. The relationships are not mutually exclusive. We consider a pair to have a given relationship if $i$ has declared that relationship with $j$ in at least half of the six surveys during that time period (reports were given approximately every month and a half). We allow these relationships to be directed and not just reciprocal. For example, if $i$ calls, sends messages to and meets with $j$ (full connectivity in the multiplex), and considers him a close friend (maximal social relationship), however $j$ does not reciprocate the relationship or communication (minimal connectivity and social relationship reported), $j$ is still considered as a close friend of $i$ according to our definition because $i$ treats him as such.

If we denote the set of reported close friendships edges as $CF$, the set of those who reported socializing as $SC$, and the set of Facebook friendships as $FB$, the relationship between the three can be described as $CF\subset SC \subset FB$. This signifies that all close friends socialize and are Facebook friends, but not all Facebook friends socialize and are close friends.We assign the highest subset (most inclusive) set to a pair, so pairs have a single definitive social relationship for the purpose of our analysis. Overall, we have 2,179 directed pairs who have not declared any social relationship; 1,299 who are only friends on Facebook but do not socialize regularly; 586 who do socialize twice per week but are not close friends, and 462 close friends. If we were to split relationships according to online and offline presence, we can state that all social relationships have online presence in the form of Facebook friendship (all 2,347), while all non-declared social relationships do not (all 2,179). 

\begin{table*}[t!]
\footnotesize
\caption{Survey parameters and summary over time. We consider three types of summary over time: $t_{max}$ when the final reported value is taken in the final survey of the study, $t_{avg}$ is the average of all reported values and $actual$ is when the actual value is reported.}
\centering
\begin{tabular}{ l l c c c c }
\hline
category & parameters & range & summary & $avg$ & $SD$\\
\hline \hline
\multirow{2}{*}{political} & interest in politics & 0-3 & $t_{max}$ & 1.67 & 1.00\\
& political orientation & 1-7 & $t_{max}$ & 5.34 & 1.31\\ \hline
\multirow{6}{*}{health} & weight(lb) & min 81.00 - max 330.00 & $t_{avg}$ & 157.5 & 41.34\\
&height(in) & min 60.00 - max 81.00 & $t_{avg}$ & 67.4 & 4.14\\
&salads per week & min 0.00 - max 6.00 & $t_{avg}$ & 1.46 & 1.43\\
&fruits per day & min 0.00 - max 7.00 & $t_{avg}$ & 2.12 & 1.45\\
&aerobics per week(days)& min 0.00 - max 7.00 & $t_{avg}$ & 1.91& 1.9\\
&sports per week(days) & min 0.00 - max 6.00 & $t_{avg}$ & 0.89 & 1.5\\ \hline
\multirow{11}{*}{music}& indie/alternative rock & 0-3 & $t_{max}$ & 1.75 & 1.17\\
& techno/lounge/electronic & 0-3 & $t_{max}$ & 1.34 & 1.09\\
& heavy metal/hardcore & 0-3 & $t_{max}$ & 1.01 & 1.1\\
& classic rock & 0-3 & $t_{max}$ & 1.84 & 1.1\\ 
& pop/top 40 & 0-3 & $t_{max}$ & 1.23 & 1.08\\ 
& hip-hop r\&b & 0-3 & $t_{max}$ & 0.75 & 0.86\\
& jazz & 0-3 & $t_{max}$ & 1.19 & 1.03\\
&classical & 0-3 & $t_{max}$ & 1.76 & 1.09\\
& country/folk & 0-3 & $t_{max}$ & 0.84 & 0.96\\
& showtunes & 0-3 & $t_{max}$ & 1.25 & 1.14\\
& other & 0-3 & $t_{max}$ & 1.25 & 1.23\\ \hline
\multirow{2}{*}{situational}&year in college& 1-5 & $actual$ & 2.5 & 1.37\\
& residential sector & 1-8 & $actual$ & 4.9 & 2.16\\
\hline
\end{tabular} 
\label{table:survey}
\end{table*}

\subsection{User Profiles}

From survey data collected periodically over the study period, we have additional information about the participants - health, political and music preferences, as well as their residential sector and year of study. We summarize the information from each category to create a composite view of a participant's attitude. Table \ref{table:survey} contains a description of this data, further elaborated upon next.\\

\paragraph{Political.} We have information about the participants' political sentiments around the 2008 presidential election. This information includes the participant's level of political interest ranging from \emph{Very interested - 3}, \emph{Somewhat interested - 2}, and \emph{Slightly interested - 1} to \emph{Not at all interested - 0}; and political orientation from \emph{Extremely liberal - 7}, \emph{Liberal - 6}, \emph{Slightly liberal - 5}, \emph{Moderate middle of the road - 4}, \emph{Slightly conservative - 3}, and \emph{Conservative - 2} to \emph{Extremely conservative - 1}. The liberal to conservative scale is fairly fine grained and could be independent from a political party. On average, participants reported they are \emph{Slightly liberal} to \emph{Liberal}. Since the political orientation can evolve over time and may be affected by the election period as was reported in~\cite{pentland2011b}, we consider the value of these parameters reported at the end of the observation period for all students, denoted by $t_{max}$. 

\paragraph{Health.} The average of the weight, height, salads and fruits, and aerobics and sports per week allowed us to build a comprehensive health profile for each user. An average ($t_{avg}$) over the survey period is used to give a single value for each attribute, and gain an understanding of the overall health habits of each student. Previous studies on this dataset found that tie formation was strongly dependent on health factors such as aerobic exercise and other campus activities \cite{pentland2011a}.

\begin{figure*}[!t]
\includegraphics[width=\textwidth]{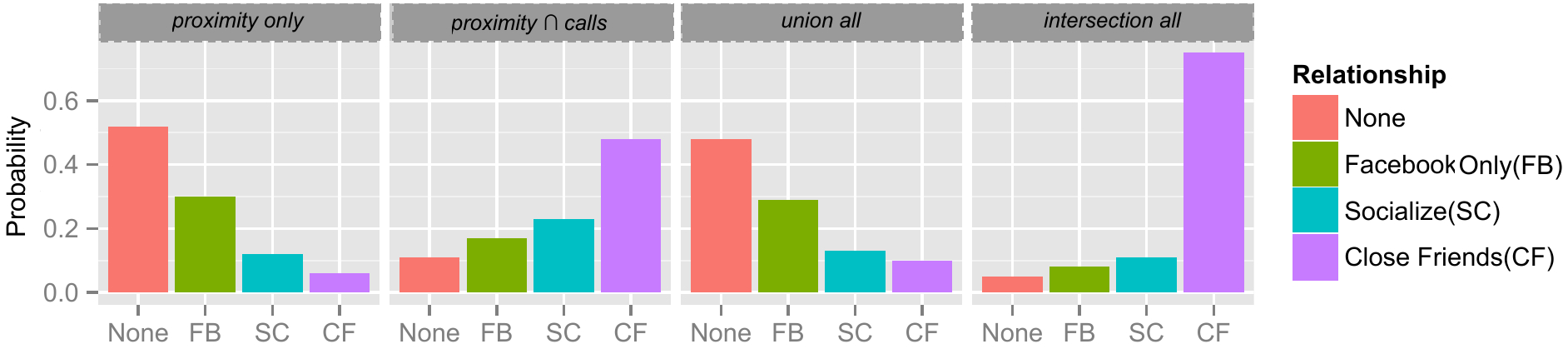}
\caption{Probability mass function $P(X=x)$ for each declared relationship $x$ within each graph aggregation. \emph{None (None)} means there was no declared relationship in the directed pair; \emph{Facebook Only (FB)} means that there was only a Facebook friendship declared between the pair; \emph{Socialize (SC)} means that the pair declared they socialize twice per week in addition to being friends on Facebook; and \emph{Close Friends (CF)} means that there was a declared close friendship in addition to all other social relationships.}
\label{fig:pmf}
\end{figure*}

\paragraph{Music.} We use the self-reported interest in each genre to build a music profile for each student. There are 11 different genres to which users have attached a preference ranging between \emph{No interest - 0}, \emph{Slight interest - 1}, \emph{Moderate interest - 2} to \emph{High interest - 3}. The most popular genre is ``classic rock" with an average rating of 1.84 and the least popular one is ``hip-hop and r\&b" with an average rating of 0.75. The association between homophily and music has been notably drawn in \cite{mark1998}, where music types were found to create niches in socio-demographic segments of society.

\paragraph{Situational.} We have information about the residential sector
(floor and building, where there are two buildings separated by a firewall only) and the college year of each student during the academic year. We coded each sector from 1 to 8 according to location and adjacency to each other. For example, sectors 1 and 2 are on the same floor, separated by a firewall, sector 3 is in the same building as sector 4 but one floor up, while sector 5 is adjacent to it and so on. In terms of year in college, students are either a \emph{Freshman - 1}, \emph{Sophomore - 2}, \emph{Junior - 3}, \emph{Senior - 4} or a \emph{Graduate Resident Tutor - 5}. These situational factors have been previously been found to be highly indicative of a social tie \cite{pentland2011a}. 

\section{Results}
We now address our hypotheses, firstly by defining the relationship between social ties and multiplexity, showing that the more communication channels are utilized, the stronger the social tie is indeed (H1). We then define a multiplex weight as a measure of tie strength, which significantly correlates with the profile similarity of students  (H2). We go on to distinguish the presence of homophily on a network level in the student community (H3). Finally, we measure whether students are potentially exposed to more diverse resources through less similar ties - online or offline, and find that although Facebook friendship is relevant to situational factors,  greater political and music diversity exposure can be found online (H4).

\subsection{H1: Multiplexity \& Social Ties}

In media multiplexity, the use of many different media indicates a strong tie~\cite{hay2005}. Here, we examine the strength of online and offline ties by considering the multiplexity of communication between students. Due to the relationships between the layers (denoted as $S$: SMS, $C$: calls, and $P$: proximity) being $S\subset C\subset P$, we have four possible non-redundant aggregations: (1) proximity layer only, which includes edges present \emph{only} on the proximity layer;
(2) $proximity \bigcap calls$, where we have those edges present on both the proximity and call layer; 
(3) $proximity \bigcup calls \bigcup SMS$, or the union of all layers; and 
(4) $proximity \bigcap calls \bigcap SMS$, or the intersection of all layers.

In Fig.~\ref{fig:pmf}, we compare the probability of each of the three social relationships (and that of having no relationship) within each of the aggregations described above. For each one, we measure the probability of having a given social tie.
Overall, we observe that single-layer communication is indicative of no relationship or an online social tie. As the number of layers increases to two, we find that the probability of having no tie decreases dramatically, while the probability of having a stronger offline tie increases. At the highest level of multiplexity, which in our case is three layers, we find the highest probability of friendship. This is aligned with previous studies of media multiplexity, and demonstrates the same principle with just a few mobile communication layers. 

Most strikingly, the highest probability of close friendship ($P=0.75$) occurs at the intersection of the three layers (\textit{intersection all} in figure). In this aggregation, all other social ties are underrepresented, highlighting the relationship between high multiplexity and strong ties (close friendship). The union and proximity only aggregations reflect the probability of having no social relationship ($P=0.5$), and also of being Facebook friends only ($P=0.3$). The intersection between the proximity and call layers on the other hand ($proximity \bigcap calls$), gives a more balanced representation of the different relationships with a 0.5 probability of close friendship, a 0.23 probability of socializing twice per week, and a 0.17 probability of being friends only on Facebook. The total probability of being friends on Facebook if two students have met during the period is defined by the total probability of a social tie, since all social ties (all relationships except ``None") are also present on Facebook. This gives a 0.5 probability of being friends online if the pair is connected on one layer (in our context the proximity layer), 0.9 if connected on two (proximity and calls), and with certainty if connected on all layers. 

Given the above observations, we can conclude that \emph{the more communication channels utilized, the stronger the tie} (H1), and that the level of multiplexity is a good indicator of tie strength.
If we consider the number of layers as an indicator of tie strength, we can describe the strength of a tie in terms of a multiplex edge weight in the network as:

\begin{equation}
mw_{ij} = \sum_{\alpha=1} ^M \frac{a_{ij}^\alpha}{M}
\end{equation}

where $M$ is the total number of layers in the multiplex. For example, if two students ($i$ and $j$) utilize all possible channels for communication (in our case $M = 3$), $mw_{ij}$ will be equal to 1, whereas if they use one channel, $mw_{ij}$ will be 1/3 . Next, we will apply this multiplex weight to assess the relationship between multiplexity and user profile similarity in terms of political, music, health and situational factors.

\subsection{H2: Multiplexity \& Profile Similarity}

With the presumption that individuals with stronger (multiplex) ties bear greater similarity, we compare the similarity between student profiles to the strength of their multiplex relationship as defined by the multiplex weight ($mw$).
We derive the similarity scores between students, using the cosine similarity of the vector of attributes for each category - music, health, political, and situational for each pair of students. These values are described in detail in Table~\ref{table:survey}. As an example, if two students are both \emph{Somewhat interested} in politics ($value$ = 2), and one is \emph{Slightly liberal} ($value$ = 5), while the other is \emph{Slightly conservative} ($value$ = 3), their cosine similarity ($sim = 0.98$) would be higher than a pair of students with the same political orientations but where one student is \emph{Not at all interested} ($value$ = 0) and the other is \emph{Very interested} ($value$ = 3, $sim$ = 0.7). Each category has a different number and range of attributes, and the similarity scores vary accordingly, however the magnitude is consistent and allows us to perform graph correlation analysis, as described next.

To find the relationship between the multiplexity ($mw_{ij}$) and the profile similarity of two individuals,  we use a standard matrix correlation coefficient. Given two generic graphs represented by the $N$x$N$ weighted adjacency matrices $A^a$ and $A^b$, we first define the correlation coefficient per node $C_i$ as follows:

\begin{equation}
C_i = 
\frac{\sum\limits_{j=1}^N w_{i,j}^aw_{i,j}^b}
{\sqrt{\sum\limits_{j=1}^N w_{i,j}^a\sum\limits_{j=1}^N w_{i,j}^b}}
\end{equation}

From the definition of the correlation coefficient per node, we can derive the graph correlation coefficient, which measures the correlation between the two weighted matrices as follows:
\begin{equation}
C(A^a, A^b) = \frac{\sum\limits_i^N C_i}{N}
\end{equation}

We calculate the graph correlations between the adjacency matrix of the multiplex and that of the pairwise similarity per  category. In essence, we compare each $mw_{ij}$ with its correspondent similarity weight, and then take the average for the whole graph (see Table \ref{table:corr}). We find that there is a significant positive relationship between the multiplex edge weights and the similarity across categories.

\begin{table}[h!]
\caption{Graph correlations between multiplex weight and each similarity score, p-value $<$ \emph{0.001 ’**’, 0.01 ’*’}.}
\small
\centering
\begin{tabular}{l l l l l }
\hline \hline
political & music & health & situational\\
\hline
0.6** & 0.49* & 0.6** & 0.56*\\
\hline
\end{tabular}
\label{table:corr}
\end{table}

The highest correlations are with the political and health factors ($C=0.6$ for both), signifying that these are most closely related to the multiplex tie strength but comparably so are also situational factors ($C=0.56$), and music ($C=0.49$). This means that multiplex ties tend to be observed in conjunction with high profile similarity. 

\begin{figure}[t!]
\centering 
\subfloat[political]
{\label{fig:rank1}\includegraphics[scale=0.45]{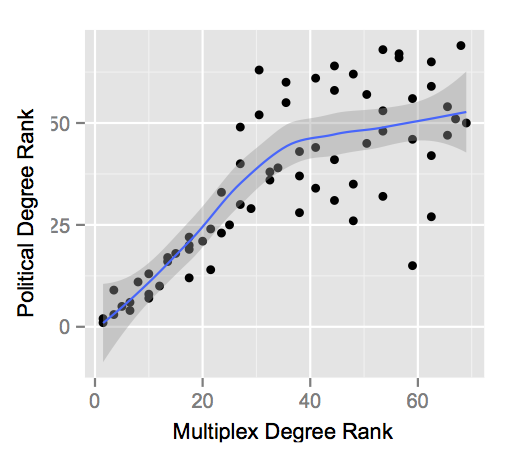}}
\subfloat[health]
{\label{fig:rank2}\includegraphics[scale=0.45]{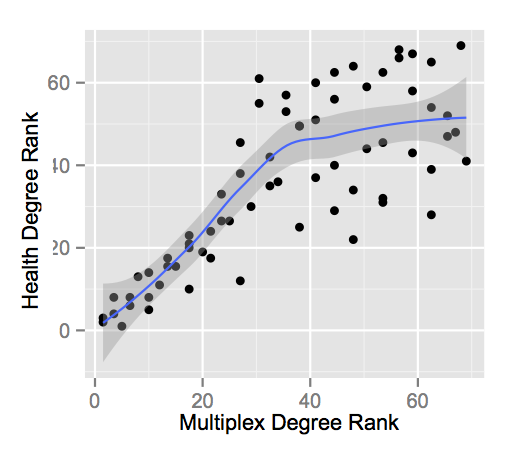}}
\\
\subfloat[music]
{\label{fig:rank3}\includegraphics[scale=0.45]{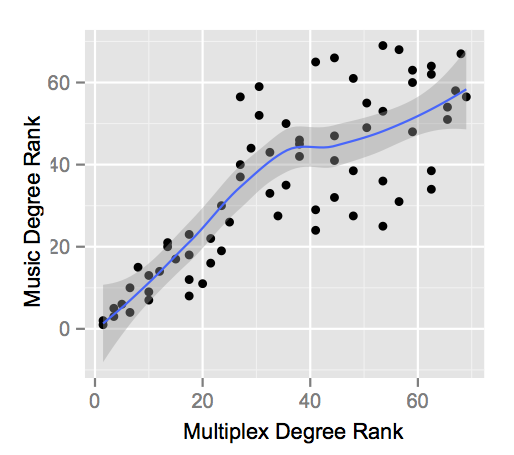}}
\subfloat[situational (year/floor)]
{\label{fig:rank4}\includegraphics[scale=0.45]{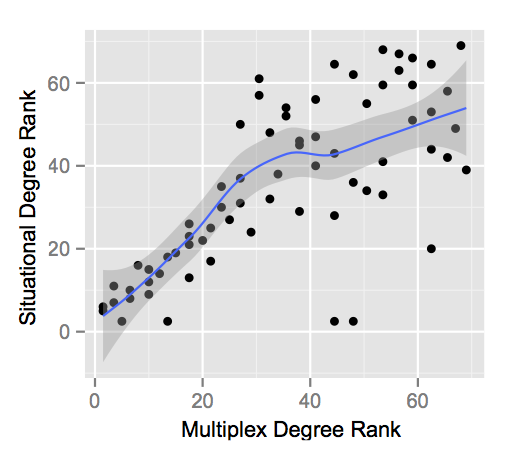}}
\caption{Spearman degree rank correlations ($\rho$) between similarity and multiplex networks. (a) political $\rho = 0.78$ (b) health $\rho = 0.79$ (c) music $\rho = 0.81$ (d) situational $\rho = 0.73$, all p-values $<$ 0.01.}
\label{fig:ranks}
\end{figure}

Next, we measure the Spearman rank degree correlations between the weighted multiplex degree and weighted similarity degree of each node and observe the effects on a neighborhood  level, displayed in Fig.~\ref{fig:ranks}. Those nodes with a high similarity degree also have a high multiplexity degree in consistence with the graph correlations on a per edge basis. This signifies that students who have more multiplex ties are also more similar to their neighbors than less popular ones in terms of degree rank.

From the correlations on a per edge level, along with correlations on the neighborhood level, we can confirm that \emph{the greater the multiplexity, the greater the similarity observed across categories} (H2). Homophily, which is observed at the network level, is explored in the following section.

\subsection{H3: Multiplexity \& Homophily}

Homophily is a network phenomenon, which is distinct from homogeneity in that it implies the occurrence of non-random similarity in pairs of connected nodes. 
We measure the presence or absence of homophily in the student community as a function of network distance and profile similarity. 
We define distance, as the standard weighted network distance. We weight our network using the multiplex weight $mw_{ij}$. The distance is then equivalent to the shortest path between two nodes in the network. A distance of 0 is indicative of full connectivity in the multiplex. This means that the pair is connected on all three layers. Direct connectivity of one hop exists up to a multiplex weight of 0.2, where weights are normalized in the range 0-1. A distance of 1 represents a non-existent path between two nodes. 

\begin{figure}[t!]
\centering 
\subfloat[political]
{\label{fig:hom1}\includegraphics[scale=0.5]{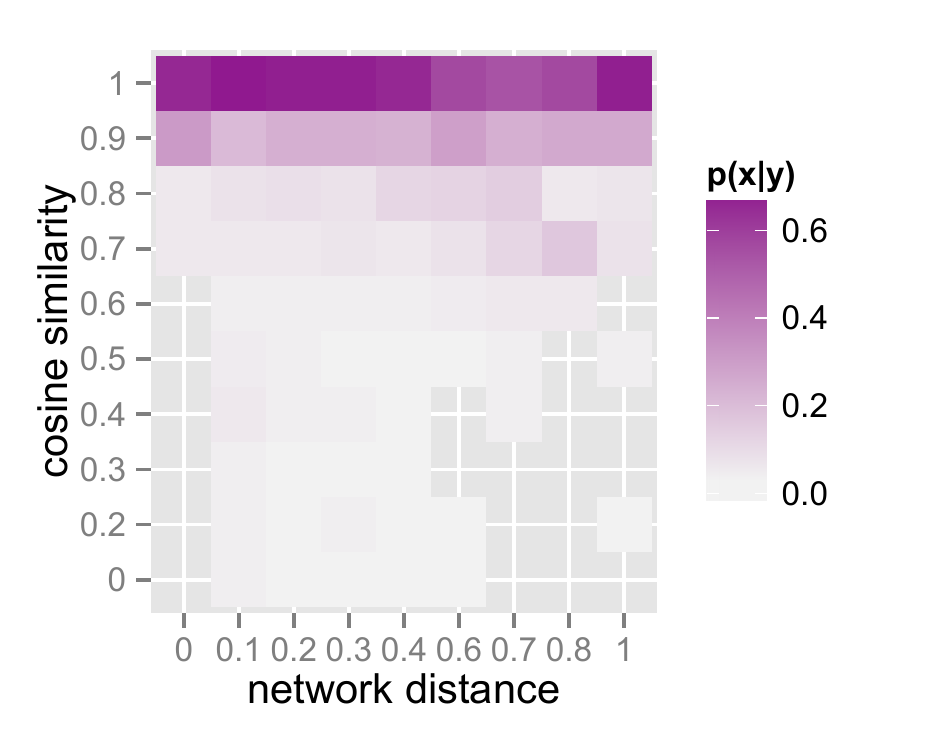}}
\subfloat[health]
{\label{fig:hom2}\includegraphics[scale=0.5]{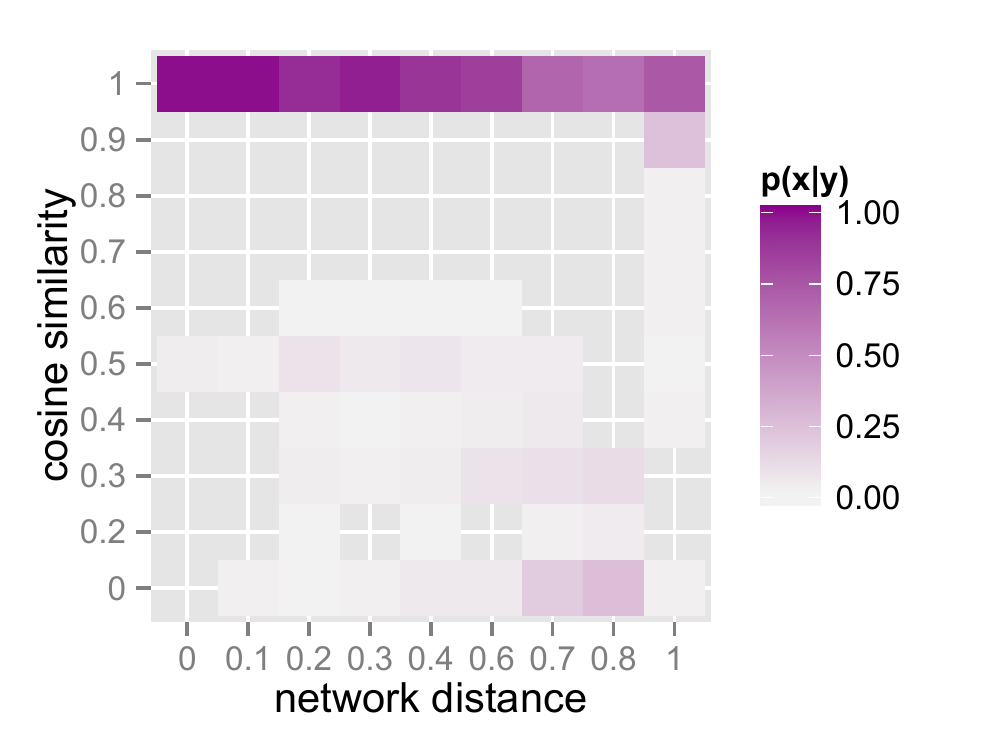}}\\
\subfloat[music]
{\label{fig:hom3}\includegraphics[scale=0.5]{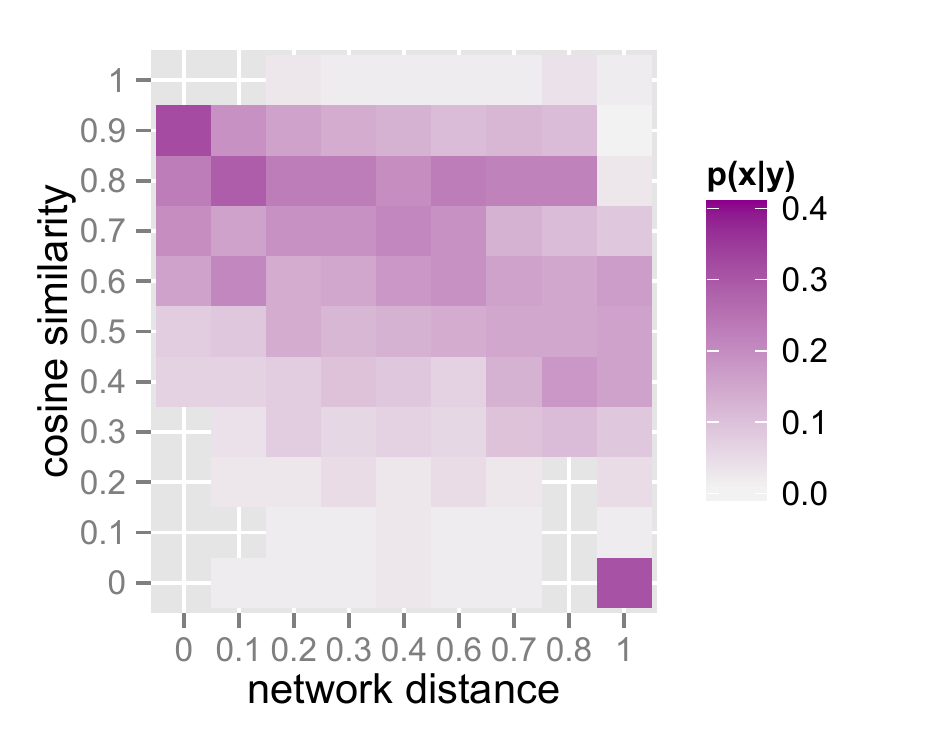}}
\subfloat[situational (year/floor)]
{\label{fig:hom4}\includegraphics[scale=0.45]{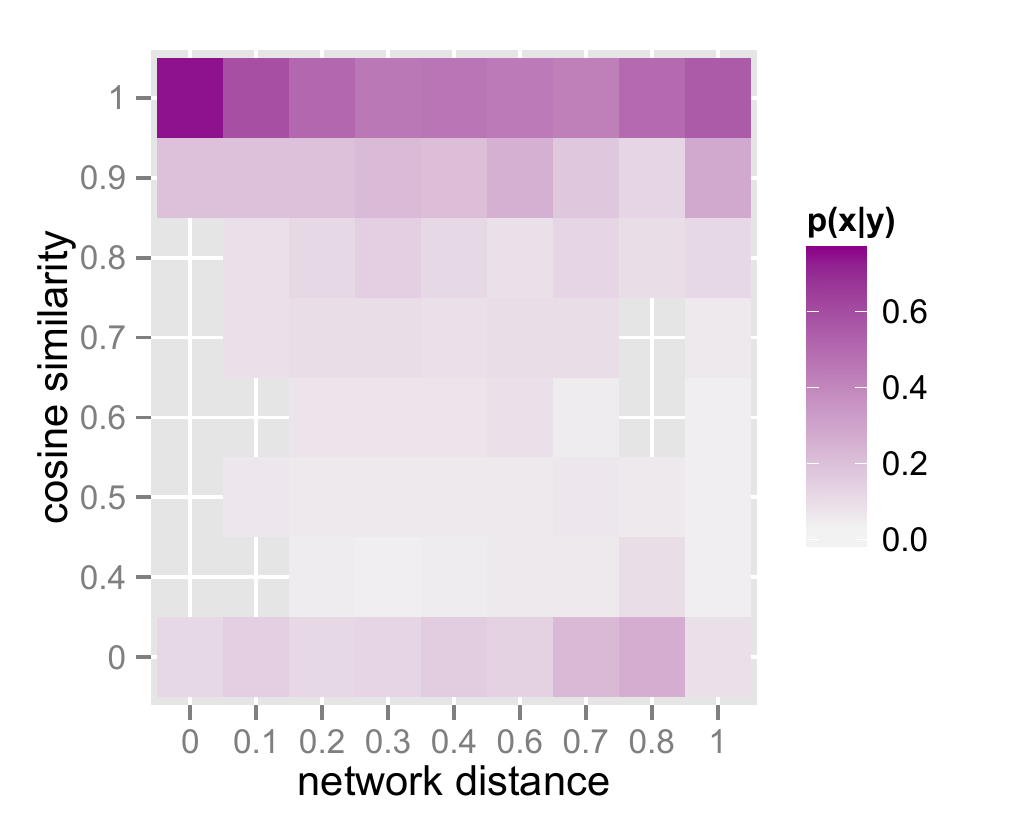}}
\caption{Conditional probability of similarity given a certain network distance ($P(x|y)$) in terms of multiplex weighted shortest path. We consider the entire multiplex networks, where the maximum distance was 0.9, and 1 is where no path exists between two nodes.}
\label{fig:sims}
\end{figure}

With the expectation that individuals at a shorter distance in the multiplex network are more similar than those further away, we measure the conditional probability of the similarity between pairs of connected students given a specific network distance, as shown in
Fig.~\ref{fig:sims}. At first glance, we can distinguish between the top graphs (\ref{fig:hom1} and \ref{fig:hom2}) as having consistently high probability of high similarity over distance, and the bottom graphs (\ref{fig:hom3} and \ref{fig:hom4}) as having a diagonal distribution of high probability, with high similarity probability decreasing over distance. 

The first two figures show an overall homogeneity in terms of political and health factors. At a distance of 1 (non-connected nodes), the probability of high similarity is still high, indicating that two nodes with high multiplexity and high similarity in these categories could be connected at random. High homogeneity can be expected in the study, given that students are co-residing and share the same context.

On the other hand, homophily exists where there is non-random similarity between individuals with shorter multiplex distance. This is most evident in subfigure~\ref{fig:hom3} - music preferences, where there is a clear shift in high probability from top left to low right as distance increases. This means that those pairs at a short distance in the network, are also highly likely to have a similar taste in music ($sim=0.9$, where $dist=0$), whereas those pairs who are further from each other in the network have a lower similarity in music taste. For unconnected pairs, the similarity is especially low (near 0), indicating that edges in the network with respect to music are non-random and highly dependent on musical preferences. 

Fig.~\ref{fig:hom4} on the other hand shows an interesting divide between low and high similarities. Most pairs are grouped into very high or very low similarity, and appear at the top row and bottom row of the graph. Therefore, students tend to be either in the same year and floor or in different years and different floors, which may be as a result of room allocation according to year. There is a high chance ($P=0.7$) that those who live and study together ($sim=1$), also have a highly multiplex tie, represented by the distinguishable top left tile at position (0,1), and less so for those who live further apart and/or are in a different year. 

Despite the community being highly homogeneous in political and health aspects, we find that \emph{the shorter the distance in the multiplex network, the greater the similarity between two nodes} (H3) in the music and situational categories, which is indicative of homophily. The decrease in diversity of resources such as information is a potentially harmful effect of homophily and homogeneity alike. We therefore conclude our analysis by exploring the role of online and offline social circles in diversifying information in the student community in the following section.

\subsection{H4: Diversity in Offline \& Online Social Circles}

So far, we have identified homogeneity and homophily within the student community. Especially during the formative university years, it is important to maintain diverse information and resources. In this section, we assess the value of social media such as Facebook in introducing  diversity across the categories taken into consideration in this work by comparing students' offline social circles with their online ones. 

We define diversity as the opposite of homogeneity, and measure the $\Delta$ difference (change) in similarity between the online and offline neighborhoods of each node. The online neighborhood of a node includes all nodes to which it is connected where there is any declared social relationship, since we know that all relationships are a subset of Facebook ($CF\subset SC \subset FB$).
The offline neighborhood includes all other connections to nodes in the graph, where there is no stated social relationship in the surveys.
To obtain the $\Delta$ change between the online and offline similarity, we subtract the average online similarity of an ego from the average offline similarity. Formally:

\begin{equation}
\Delta sim = \frac{\sum sim_{\text{\emph{off}}}}{|N_{\text{\emph{off}}}|} - \frac{\sum sim_{on}}{|N_{on}|}
\end{equation} 

where $sim_{\text{\emph{off}}}$ and $sim_{on}$ are the similarity values between the offline and online contacts of a given student, and $N_{\text{\emph{off}}}$ and $N_{on}$ are his offline and online neighborhoods. 

\begin{figure}[h!]
\centering
\includegraphics[scale=0.8]{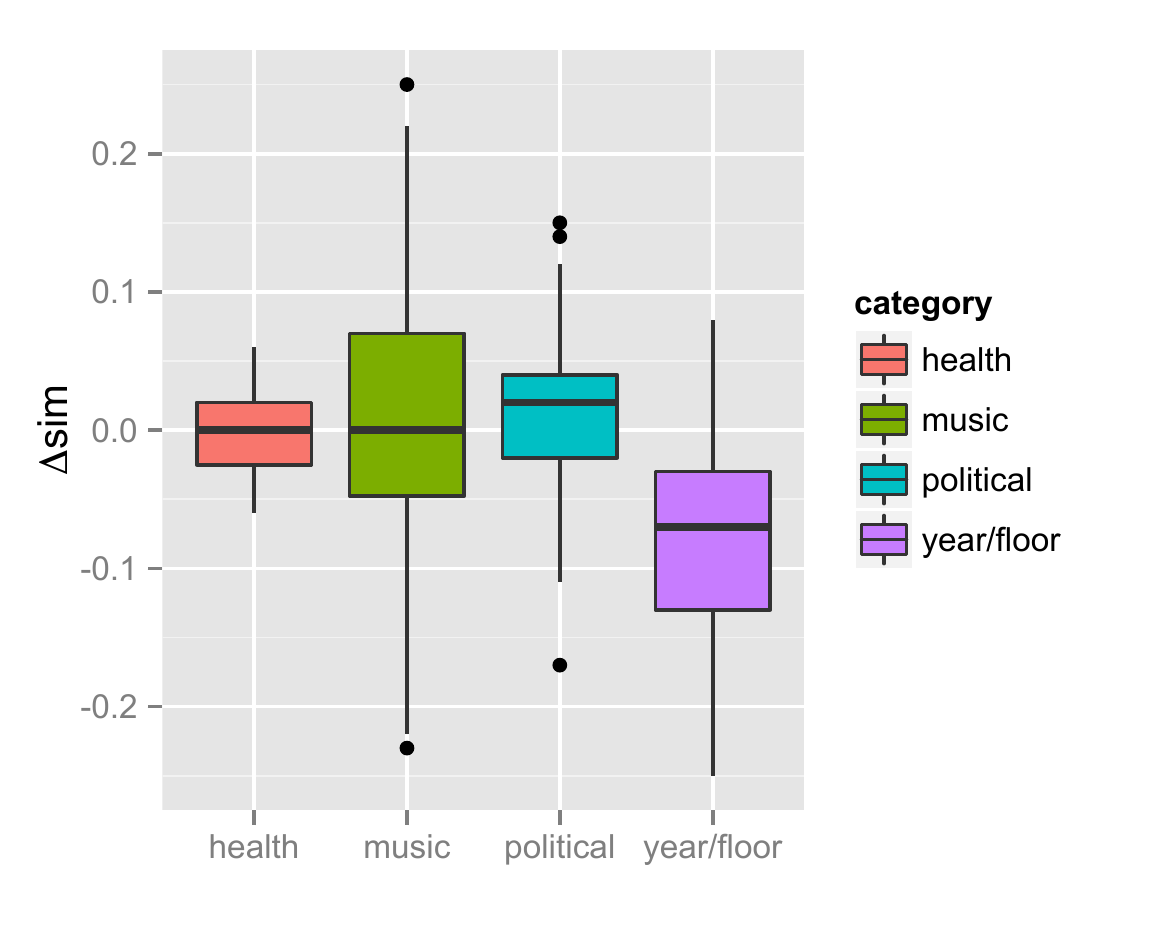}
\caption{Box-and-whiskers plot of the difference between online and offline average neighborhood similarity.}
\label{fig:diff}
\end{figure}

In Fig.~\ref{fig:diff}, values above 0 indicate that there is greater online diversity than offline, whereas values below zero indicate that there is a lower diversity online than offline. The boxplot can be read as the distribution of $\Delta sim$, where the values are divided into quartiles. The top and bottom continuous lines (or whiskers) represent the top and bottom quartiles, whereas the values inside the box, below and above the thick line representing the median, are the upper and lower mid-quartiles. 

We can see immediately that the situational (year/floor) category online is largely less diverse than offline. In other words, students are more often Facebook friends with other students in their year and residential sector but may meet offline with students from other residential sectors and/or years. Interestingly, we find that in terms of political orientation, Facebook friendship between students tends to be more varied than offline, with more than 50\% of values above zero. In terms of music preference and health, we have a 50/50 distribution, where half of online relationships are more varied than offline and vice versa. However, the music distribution is more wide spread, meaning that there is a big disparity between offline and online neighborhoods. On the other hand, health indicators are more clustered around the median, indicating small disparities between online and offline neighborhood diversity. 

From this analysis, the situational (year/floor) category appears to be the most important factor for Facebook friendship, however, we observe that \emph{greater diversity is possible online in terms of politics and music}. Music exhibits the most observable homophily in Fig.~\ref{fig:sims}, yet some students have more musically diverse Facebook friends than offline contacts. This is due to the steep decline in similarity in Fig.~\ref{fig:hom3}, after a distance of 0, which means that although one is very likely to have music in common with their close friends, the people they socialize with or have just as Facebook friends can be highly diverse. 


\section{Discussion \& Conclusions}

\paragraph{Summary of Contributions.}

This work makes several contributions to the study of human interactions online and offline. We firstly describe a generalizable model for defining multiplex tie strength through different media. This model is grounded on previous work on multiplexity~\cite{haythornthwaite1998}, and attempts to fill a gap in homophily research, identified in~\cite{mcpherson2001} by advancing the understanding of multiplex social ties, social distance, and applying it to the online and offline context as required by the two-dimensional nature of human relations nowadays. 

We additionally show in the context of our study that in social circles such as those formed in a college campus, strong ties with high multiplexity bear greater similarity, and highly ranked nodes have many connections with high similarity. We find more similar nodes are at a shorter distance in the network in situational and music aspects, as evidence of homophily, while political and health factors indicate high homogeneity in the community. We also find that for three frequently shared categories of information online - politics, music, and health, users were part of more or equally diverse social circles online, however their online connectivity is highly dependent on situational factors such as year in college and residential sector. 

\paragraph{Limitations.}

The academic environment has been a rich source of sociological studies, and education itself is a strong factor in homophily~\cite{mcpherson2001}. Studying homophily in the college context is relevant to the socio-demographic background in the institution, and as we have observed in the present work, consists of a highly homogeneous group of people. Unfortunately, parallel online and offline data is rarely available and similar studies would benefit in the future from wider-context at a larger scale. Nevertheless, it is precisely in such homogeneous communities that diversity of information is most needed, especially in the formative university years. 

Despite the potential for exposure to greater information diversity online, we are not able to conclude as to the usage patterns and content of the students' newsfeeds on Facebook. There are many additional constraints to receiving diverse information such as time and attention spent on Facebook, aggregation of posts, and blocking of content by the user - not considered in the present work.  
Nevertheless, we are able to observe a satisfactory measurement of multiplex tie strength, and present a novel model generalizable to all communication interactions.

\paragraph{Implications.}

Our work reflects the role of social media in maintaining diverse social circles, highlighting a challenge for the design of such services - allowing for diversity while maintaining relevance in an increasingly populated online information stream. This has become of recent public concern, as the term \emph{filter bubble} emerged to describe the personalization of services such as Facebook and Google as an ``echo chamber" of censorship in search and information aggregation~\cite{eli2011filter}. The concern is that receiving only agreeable and relevant information may obscure objectivity and limit access to potentially useful and diverse resources. 

Recent work, however,  has demonstrated that Facebook users are more likely to share diverse content from weaker ties than from stronger, more similar ties as a result of generally having more weak than strong ties online~\cite{bakshy2012}. On the other hand, diversity online is relative to diversity offline - we tend to associate ourselves online with people we meet offline or who share similar interests and likely come from a similar background. In this work, student ties at the same educational institution were strongly affected by residential sector and year in college but exhibited diversity in other aspects online. 

It is evident that online social circles are maintained at a lower time cost and social media such as Facebook have become an important component of human communication. Facebook not only helps decrease tie decay over distance and time~\cite{dunbar2011} but can also enhance tie strength through online interaction~\cite{burke2014} and instill a sense of closeness and stability to geographically distant and communication-restricted ties~\cite{vitak14}. Facebook serves ties of varying strength, as reflected in the present work, and allows us to stay engaged with a large subset of the people we meet offline, stretching the boundaries of our social circles. 
Increases in audience size and diversity on Facebook are associated with an increase in the amount of broadcasted information online~\cite{vitak12}. The passive consumption of broadcasted information from friends on Facebook can also lead to an increase in tie strength~\cite{burke2014}, and therefore diverse and large audiences on Facebook contribute to the overall increase of social capital, as is also implied by the present work. These technology effects have much more profound implications outside the geographically constrained college environment, nevertheless, even small increments in the diversity of homogeneous social circles can be of great benefit and may also be encouraged by design. 

When it comes to the depth of human relations, it is difficult to reduce them to any single value. Our work goes one step in the direction of applying more dimensionality to social ties, a direction which will hopefully bring greater understanding to the online and offline aspects of human social life and their interdependence.\\

{\small {\bf Acknowledgements}. The authors would like to thank Pietro Panzarasa for the insightful comments. This work was supported by the Project LASAGNE, Contract No. 318132 (STREP), funded by the European Commission.}

\bibliographystyle{aaai}
{\scriptsize
\bibliography{biblio}}

\end{document}